\documentclass[11pt]{article}

\usepackage{PRIMEarxiv}

\usepackage[utf8]{inputenc} 
\usepackage[T1]{fontenc}    
\usepackage{hyperref}       
\usepackage{url}            
\usepackage{booktabs}       
\usepackage{amsfonts}       
\usepackage{nicefrac}       
\usepackage{microtype}      
\usepackage{fancyhdr}       
\usepackage{graphicx}       
\usepackage{lipsum}
\graphicspath{{media/}}     

\usepackage{amsmath, amssymb} 
\usepackage{subcaption}
\usepackage{float}
\usepackage{adjustbox}        
\usepackage[authoryear]{natbib} 
\usepackage{rotating}         
\usepackage{caption}          
\title{Comparative Analysis of Novel NIRMAL Optimizer Against Adam and SGD with Momentum}

\author{
  Nirmal Gaud \\
  CEO and Master Trainer \\
  ThinkAI - A Machine Learning Community \\
  \texttt{nirmal.gaud.ai@gmail.com} \\
  \And
  Surej Mouli \\
  Lecturer \\
  Aston University, Birmingham, UK \\
  surej@live.co.uk \\
  \And
  Preeti Katiyar \\
  Assistant Professor \\
  Delhi Technical Campus Greater Noida \\
  \texttt{katiyar.preeti29@gmail.com} \\
  \And
  Vaduguru Venkata Ramya \\
  Teaching Assistant \\
  Sri Balaji University Pune, India \\
  \texttt{ramyavvphd@gmail.com} \\
}

\begin{document}
\maketitle

\begin{abstract}
{This study proposes NIRMAL (Novel Integrated Robust Multi-Adaptation Learning), a novel optimization algorithm that combines multiple strategies inspired by the movements of the chess piece. These strategies include gradient descent, momentum, stochastic perturbations, adaptive learning rates, and non-linear transformations. We carefully evaluated NIRMAL against two widely used and successful optimizers, Adam and SGD with Momentum, on four benchmark image classification datasets: MNIST, FashionMNIST, CIFAR-10, and CIFAR-100. The custom convolutional neural network (CNN) architecture is applied on each dataset. The experimental results show that NIRMAL achieves competitive performance, particularly on the more challenging CIFAR-100 dataset, where it achieved a test accuracy of 45.32\%and a weighted F1-score of 0.4328. This performance surpasses Adam (41.79\% accuracy, 0.3964 F1-score) and closely matches SGD with Momentum (46.97\% accuracy, 0.4531 F1-score). Also, NIRMAL exhibits robust convergence and strong generalization capabilities, especially on complex datasets, as evidenced by stable training results in loss and accuracy curves. These findings underscore NIRMAL's significant ability as a versatile and effective optimizer for various deep learning tasks.}
\end{abstract}


\section{Introduction}
Deep learning models are highly dependent on efficient optimization algorithms to minimize losses and improve generalization scores. This becomes more important when performing complex tasks, such as image classification. Although Stochastic Gradient Descent (SGD) with Momentum and Adam are widely used fundamental optimizers in the field, they can present challenges such as slow convergence rates or high sensitivity to hyperparameter tuning \citep{kingma2017adam, liu2020sgd, acharya2015sgd}. To address these issues and explore the novel optimization algorithm, we propose NIRMAL (Novel Integrated Robust Multi-Adaptation Learning), a hybrid optimizer that strategically combines multiple distinct optimization strategies, conceptually inspired by the diverse movements of chess pieces.

This paper presents a comprehensive comparative analysis of NIRMAL against the famous Adam and SGD with Momentum optimizers. For this purpose we used four prominent image classification datasets: MNIST, FashionMNIST, CIFAR-10, and CIFAR-100. The selection of these datasets, ranging from relatively simple (MNIST) to highly complex (CIFAR-100), facilitates a thorough and robust evaluation of each optimizer's performance under varying data complexities. We used custom CNN for each dataset to ensure a suitable architectural fit. Performance assessment is performed using key metrics that include test accuracy, test loss, and weighted F1-score, complemented by training history analyzes and detailed classification reports.

The primary objectives of this research are as follows.
\begin{enumerate}
    \item Detail the formulation and underlying principles of the NIRMAL optimizer.
    \item Conduct a comparative evaluation of NIRMAL's performance against two leading optimizers (Adam and SGD with Momentum) on standard image classification benchmarks.
    \item Analyze NIRMAL's effectiveness in terms of convergence behavior, stability during training, and generalization capabilities in different dataset complexities.
\end{enumerate}
This work aims to contribute to the development of more, efficient and effective optimization algorithms in deep learning techniques.

\section{Literature Review}
Stochastic gradient-based optimization methods are widely used in machine learning to train models efficiently on large datasets. The study in \citep{acharya2015sgd} compares different optimization techniques such as SGD, mini-batch gradient descent, and adaptive optimizers using large least squares problems. It was found that mini-batch gradient descent provides a good balance between the noisy updates of standard SGD and the stable updates of full-batch gradient descent. The study also observed that Adam performed best among adaptive optimizers, while Nesterov gave the best results among fixed learning rate methods. It suggested that adaptive optimizers work better on large datasets but may require more computation.

Another study \citep{kingma2017adam} introduced Adam, a popular optimizer that combines ideas from AdaGrad and RMSProp. Adam adapts the learning rate for each parameter by computing exponentially decaying averages of past gradients (first moment) and squared gradients (second moment) \citep{saud2023bert}. It is easy to use, memory-efficient, and works well on large-scale and high-dimensional problems. The authors provided both theoretical analysis and experimental results showing that Adam works well on noisy and non-stationary objectives. They also proposed a variant called AdaMax, which uses a different type of normalization.

The work in \citep{liu2020sgd} focuses on SGD with Momentum (SGDM). Although SGDM is commonly used in practice, its theoretical understanding was limited. This study shows that SGDM can achieve the same convergence rate as SGD in both strongly convex and non-convex settings. It also introduced multistage SGDM, where learning rates and momentum values are changed in stages during training. The authors proved that this approach improves performance and gave theoretical support for a widely used training strategy. They also suggested further research to see whether SGDM can outperform SGD in specific cases.

\section{Optimization Algorithms}
This section provides a detailed description of the three optimization algorithms evaluated in this study: NIRMAL, Adam, and SGD with Momentum. Their mathematical formulations are presented to highlight their unique update mechanisms.

\subsection{NIRMAL Optimizer}
NIRMAL's innovative design integrates five distinct strategies, each symbolically associated with a chess piece, contributing a specific aspect to the parameter update:
\begin{itemize}
    \item \textbf{Wazir (Gradient Descent)}: Represents the direct application of the gradient for immediate parameter updates.
    \item \textbf{Elephant (Momentum)}: Incorporates a velocity term to accelerate convergence in consistent directions, smoothing oscillations.
    \item \textbf{Knight (Stochastic Perturbations)}: Introduces random noise to help escape shallow local minima and explore the loss landscape more effectively.
    \item \textbf{Camel (Adaptive Learning Rate Scaling)}: Dynamically adjusts the learning rate based on historical gradient magnitudes, ensuring robust step sizes.
    \item \textbf{Horse (Non-linear Momentum Transformation)}: Applies a non-linear transformation to the momentum term, potentially enhancing stability and controlling step magnitudes.
\end{itemize}
For a parameter \(\theta\) with gradient \(g_t = \nabla_\theta L(\theta_t)\) at time step \(t\), NIRMAL's update rule is formulated as follows:
\begin{align}
    m_t &= \mu m_{t-1} + (1 - \mu) g_t, \label{eq:nirmal_momentum} \\
    v_t &= \beta v_{t-1} + (1 - \beta) g_t^2, \label{eq:nirmal_variance} \\
    \Delta_{\text{wazir}} &= -\eta g_t, \label{eq:nirmal_wazir} \\
    \Delta_{\text{elephant}} &= -\eta m_t, \label{eq:nirmal_elephant} \\
    \Delta_{\text{knight}} &= \eta \kappa \mathcal{N}(0, 1), \label{eq:nirmal_knight} \\
    \begin{split}
    \Delta_{\text{camel}} &= -\eta \gamma \frac{m_t}{\sqrt{v_t} + \epsilon}, \label{eq:nirmal_camel}
    \end{split} \\
    \Delta_{\text{horse}} &= -\eta \lambda \tanh(m_t), \label{eq:nirmal_horse} \\
    \begin{split}
    \Delta_{\text{total}} &= w_{\text{wazir}} \Delta_{\text{wazir}} + w_{\text{elephant}} \Delta_{\text{elephant}} + w_{\text{knight}} \Delta_{\text{knight}} \\
    &\quad + w_{\text{camel}} \Delta_{\text{camel}} + w_{\text{horse}} \Delta_{\text{horse}}, \label{eq:nirmal_total}
    \end{split} \\
    \theta_{t+1} &= \theta_t + \Delta_{\text{total}}. \label{eq:nirmal_update}
\end{align}
The hyperparameters used for NIRMAL are: learning rate \(\eta = 10^{-3}\), momentum decay rates \(\mu = 0.9\) and \(\beta = 0.999\), a small constant \(\epsilon = 10^{-8}\) to prevent division by zero, stochastic perturbation scale \(\kappa = 0.01\), adaptive scaling factor \(\gamma = 1.5\), and non-linear transformation scale \(\lambda = 0.5\). The contribution of each component to the total update is controlled by weights: \(w_{\text{wazir}} = 0.3\), \(w_{\text{elephant}} = 0.25\), \(w_{\text{knight}} = 0.1\), \(w_{\text{camel}} = 0.2\), and \(w_{\text{horse}} = 0.15\). Furthermore, the weight decay (\(\alpha\)) is incorporated by modifying the gradient as \(g_t \gets g_t + \alpha \theta_t\).

\subsection{Adam Optimizer}
\textbf{Adam (Adaptive Moment Estimation)} computes adaptive learning rates for each parameter by maintaining exponentially decaying moving averages of the past gradients (\(m_t\)) and the past squared gradients (\(v_t\)). Bias correction is applied to these moving averages to account for their initialization at zero. The update rule is given by:
\begin{align}
    m_t &= \beta_1 m_{t-1} + (1 - \beta_1) g_t, \label{eq:adam_momentum} \\
    v_t &= \beta_2 v_{t-1} + (1 - \beta_2) g_t^2, \label{eq:adam_variance} \\
    \hat{m}_t &= \frac{m_t}{1 - \beta_1^t}, \quad \hat{v}_t = \frac{v_t}{1 - \beta_2^t}, \label{eq:adam_bias_correction} \\
    \theta_{t+1} &= \theta_t - \eta \frac{\hat{m}_t}{\sqrt{\hat{v}_t} + \epsilon}. \label{eq:adam_update}
\end{align}
Standard hyperparameters for Adam are used: learning rate \(\eta = 10^{-3}\), decay rates for moment estimates \(\beta_1 = 0.9\) and \(\beta_2 = 0.999\), and a small constant \(\epsilon = 10^{-8}\).

\subsection{SGD with Momentum}
\textbf{Stochastic Gradient Descent (SGD) with Momentum} is a traditional but effective optimization algorithm. It aims to speed up the SGD in the relevant direction and dampen oscillations. This is achieved by adding a fraction of the update vector of the previous time step to the current update. The velocity term accumulates gradients over time, providing inertia \citep{soudani2019image}. Its update rules are as follows:
\begin{align}
    v_t &= \mu v_{t-1} + \eta g_t, \label{eq:sgd_momentum} \\
    \theta_{t+1} &= \theta_t - v_t. \label{eq:sgd_update}
\end{align}
The parameters used are: learning rate \(\eta = 0.01\) and momentum factor \(\mu = 0.9\). For the CIFAR datasets, a weight decay of \(\alpha = 5 \times 10^{-4}\) was applied to prevent overfitting.

\section{Experimental Setup}
Our experiments were systematically conducted on four widely used image classification datasets to ensure a comprehensive evaluation:
\begin{itemize}
    \item \textbf{MNIST}: A dataset of 70,000 grayscale images of handwritten digits (0-9), divided into 60,000 training and 10,000 test images, with 10 classes.
    \item \textbf{FashionMNIST}: A direct drop-in replacement for MNIST, consisting of 70,000 grayscale images of fashion articles (e.g., shirts, trousers), with 10 classes.
    \item \textbf{CIFAR-10}: Comprises 60,000 32x32 color images in 10 classes (e.g., airplane, car), with 50,000 training and 10,000 test images.
    \item \textbf{CIFAR-100}: Similar to CIFAR-10 but with 100 classes, each containing 600 images (500 training, 100 testing).
\end{itemize}

\subsection{Convolutional Neural Network (CNN) Architectures}
Tailored CNN architectures were designed for each dataset's complexity:
\begin{itemize}
    \item For MNIST and FashionMNIST, a simpler architecture was used, consisting of two convolutional layers followed by max-pooling, and then two fully connected layers.
    \item For CIFAR-10, the architecture was augmented to include three convolutional layers, interspersed with batch normalization and max-pooling layers, culminating in two fully connected layers.
    \item For the more complex CIFAR-100, a deeper network was employed, featuring four convolutional layers, batch normalization, max-pooling, and three fully connected layers.
\end{itemize}

\subsection{Training Parameters}
All models were trained for 10 epochs using a batch size of 64. The cross-entropy loss function was utilized for training. To improve generalization and prevent overfitting, data augmentation techniques (random cropping and horizontal flipping) were applied specifically to the CIFAR-10 and CIFAR-100 datasets.

\subsection{Evaluation Metrics}
The performance of each optimizer was evaluated using three primary metrics on the test set:
\begin{itemize}
    \item \textbf{Test Accuracy}: The percentage of correctly classified images.
    \item \textbf{Test Loss}: The value of the loss function on the test set, indicating the generalization error.
    \item \textbf{Weighted F1-score}: The harmonic mean of precision and recall, weighted by the number of true instances for each label, providing a balanced measure for multiclass classification.
\end{itemize}

\subsection{Optimizer Hyperparameters}
The NIRMAL optimizer uses multiple hyperparameters: learning rate (lr=1e-3), momentum (0.9), beta (0.999), epsilon (eps=1e-8), knight scale (0.01), camel scale (1.5), horse scale (0.5), weights for Wazir (0.3), Elephant (0.25), Knight (0.1), Camel (0.2), and Horse (0.15), and weight decay (5e-4 for CIFAR10/CIFAR100, 0 otherwise). Adam uses the learning rate (lr=1e-3), betas (0.9, 0.999), and epsilon (eps=1e-8). SGD with Momentum uses learning rate (lr=0.01), momentum (0.9), and weight decay (5e-4 for CIFAR10/CIFAR100, 0 otherwise).

\section{Results and Analysis}
This section presents and analyzes the experimental results, comparing the performance of NIRMAL, Adam, and SGD with Momentum across the four benchmark datasets. Table \ref{tab:performance} summarizes the key performance metrics, followed by detailed analyzes for each dataset, including confusion matrices and training history plots for each optimizer. A comparative metrics plot provides an overview of performance across all datasets.

\begin{table}[H]
\centering
\caption{Performance comparison of optimizers across datasets. Best performance in each metric is highlighted in bold.}
\begin{adjustbox}{width=0.95\textwidth}
\begin{tabular}{lccc|ccc|ccc}
\toprule
 & \multicolumn{3}{c}{Test Accuracy (\%)} & \multicolumn{3}{c}{Test Loss} & \multicolumn{3}{c}{F1-Score (Weighted)} \\
\cmidrule(lr){2-4} \cmidrule(lr){5-7} \cmidrule(lr){8-10}
Dataset & NIRMAL & Adam & SGD+M & NIRMAL & Adam & SGD+M & NIRMAL & Adam & SGD+M \\
\midrule
MNIST & 99.18 & \textbf{99.22} & 99.21 & 0.0272 & 0.0302 & \textbf{0.0250} & 0.9918 & \textbf{0.9922} & 0.9921 \\
FashionMNIST & \textbf{92.36} & 92.16 & 92.16 & \textbf{0.2211} & 0.2681 & 0.2264 & \textbf{0.9226} & 0.9217 & 0.9211 \\
CIFAR-10 & 75.02 & \textbf{78.95} & 78.01 & 0.7125 & \textbf{0.6149} & 0.6324 & 0.7512 & \textbf{0.7892} & 0.7787 \\
CIFAR-100 & 45.32 & 41.79 & \textbf{46.97} & 2.0516 & 2.1912 & \textbf{1.9408} & 0.4328 & 0.3964 & \textbf{0.4531} \\
\bottomrule
\end{tabular}
\end{adjustbox}
\label{tab:performance}
\end{table}

\subsection{MNIST}

\begin{figure}[H]
\centering
\begin{subfigure}{\textwidth}
    \centering
    \includegraphics[width=1.0\linewidth]{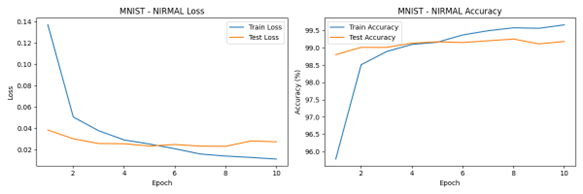}
    \caption{NIRMAL}
    \label{fig:mnist_hist_nirmal}
\end{subfigure}
\begin{subfigure}{\textwidth}
    \centering
    \includegraphics[width=1.0\linewidth]{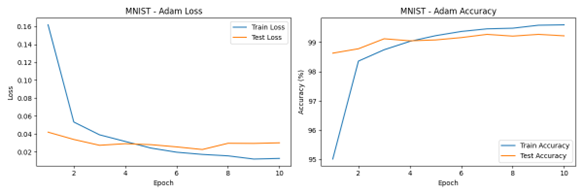}
    \caption{Adam}
    \label{fig:mnist_hist_adam}
\end{subfigure}
\begin{subfigure}{\textwidth}
    \centering
    \includegraphics[width=1.0\linewidth]{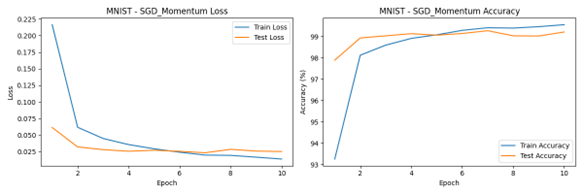}
    \caption{SGD with Momentum}
    \label{fig:mnist_hist_sgdm}
\end{subfigure}

\caption{Training history (loss and accuracy curves) for MNIST dataset across optimizers.}
\label{fig:mnist_history}
\end{figure}

On the MNIST dataset, all optimizers gave excellent results, reflecting the simplicity of this task. Adam achieved the highest test accuracy at \textbf{99.22\%}, closely followed by SGD with Momentum (99.21\%) and NIRMAL (99.18\%). SGD with Momentum recorded the lowest test loss (0.0250), indicating slightly better convergence. The training history plots (Figure \ref{fig:mnist_history}) demonstrate stable convergence across all optimizers, with NIRMAL showing minimal oscillations in loss and accuracy curves.

\subsection{FashionMNIST}

\begin{figure}[H]
\centering
\begin{subfigure}{\textwidth}
    \centering
    \includegraphics[width=1.0\linewidth]{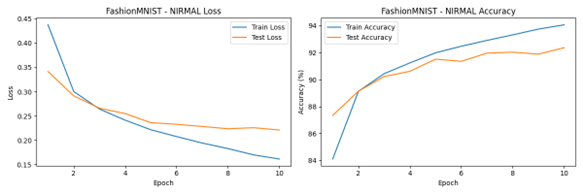}
    \caption{NIRMAL}
    \label{fig:fashionmnist_hist_nirmal}
\end{subfigure}
\begin{subfigure}{\textwidth}
    \centering
    \includegraphics[width=1.0\linewidth]{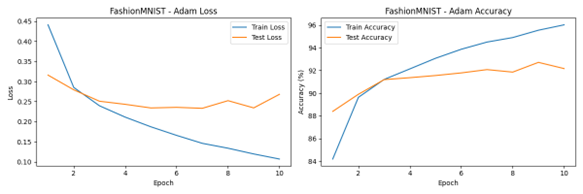}
    \caption{Adam}
    \label{fig:fashionmnist_hist_adam}
\end{subfigure}
\begin{subfigure}{\textwidth}
    \centering
    \includegraphics[width=1.0\linewidth]{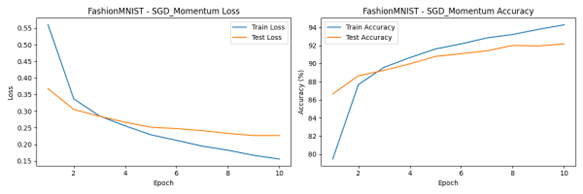}
    \caption{SGD with Momentum}
    \label{fig:fashionmnist_hist_sgdm}
\end{subfigure}

\caption{Training history (loss and accuracy curves) for FashionMNIST dataset across optimizers.}
\label{fig:fashionmnist_history}
\end{figure}

For FashionMNIST, NIRMAL demonstrated a slight edge, leading in test accuracy (\textbf{92.36\%}) and achieving the lowest test loss (\textbf{0.2211}). Adam and SGD with Momentum performed almost identically at 92.16\% accuracy. The training history plots (Figure \ref{fig:fashionmnist_history}) reveal that NIRMAL’s loss and accuracy curves are smoother, suggesting robust convergence compared to the slight fluctuations of Adam.

\subsection{CIFAR-10}

\begin{figure}[H]
\centering
\begin{subfigure}{\textwidth}
    \centering
    \includegraphics[width=1.0\linewidth]{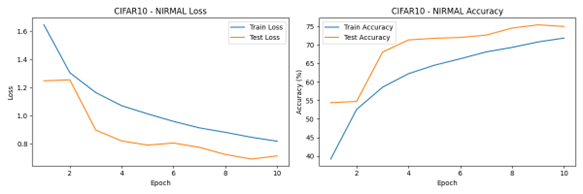}
    \caption{NIRMAL}
    \label{fig:cifar10_hist_nirmal}
\end{subfigure}
\begin{subfigure}{\textwidth}
    \centering
    \includegraphics[width=1.0\linewidth]{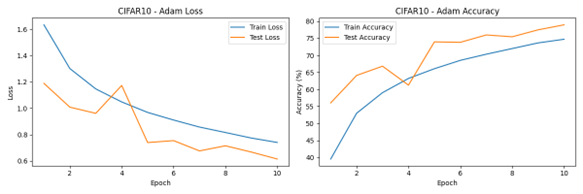}
    \caption{Adam}
    \label{fig:cifar10_hist_adam}
\end{subfigure}
\begin{subfigure}{\textwidth}
    \centering
    \includegraphics[width=1.0\linewidth]{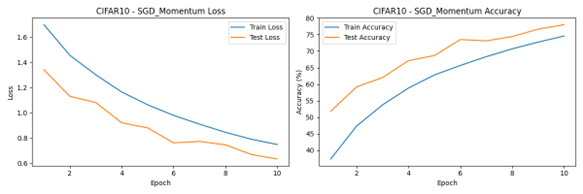}
    \caption{SGD with Momentum}
    \label{fig:cifar10_hist_sgdm}
\end{subfigure}

\caption{Training history (loss and accuracy curves) for CIFAR-10 dataset across optimizers.}
\label{fig:cifar10_history}
\end{figure}

On CIFAR-10, a more challenging dataset due to color images and complex patterns, Adam emerged as the top performer with \textbf{78.95\% accuracy} and the lowest test loss (0.6149). SGD with Momentum followed closely with 78.01\% accuracy, while NIRMAL achieved 75.02\% accuracy. The training history plots (Figure \ref{fig:cifar10_history}) indicate that Adam and SGD with Momentum converged faster, while NIRMAL’s curves show steady but slower improvement.

\subsection{CIFAR-100}
\begin{figure}[H]
\centering
\begin{subfigure}{\textwidth}
    \centering
    \includegraphics[width=1.0\linewidth]{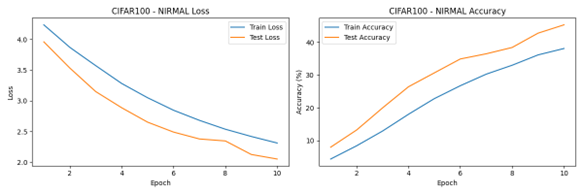}
    \caption{NIRMAL}
    \label{fig:cifar100_hist_nirmal}
\end{subfigure}
\begin{subfigure}{\textwidth}
    \centering
    \includegraphics[width=1.0\linewidth]{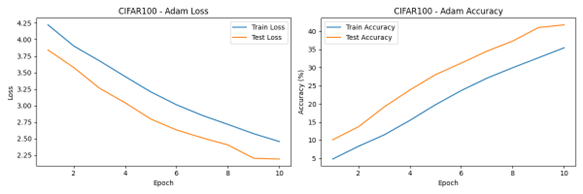}
    \caption{Adam}
    \label{fig:cifar100_hist_adam}
\end{subfigure}
\begin{subfigure}{\textwidth}
    \centering
    \includegraphics[width=1.0\linewidth]{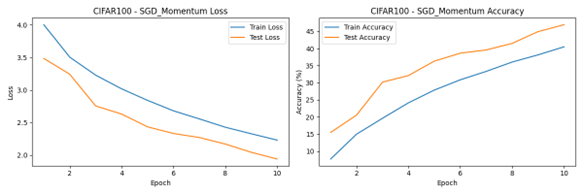}
    \caption{SGD with Momentum}
    \label{fig:cifar100_hist_sgdm}
\end{subfigure}

\caption{Training history (loss and accuracy curves) for CIFAR-100 dataset across optimizers.}
\label{fig:cifar100_history}
\end{figure}

The CIFAR-100 dataset, with its 100 classes, presented the greatest challenge. NIRMAL showed its strength, achieving a test accuracy of 45.32\% and a weighted F1-score of 0.4328, surpassing Adam (41.79\% accuracy, 0.3964 F1-score). SGD with Momentum achieved the highest accuracy (\textbf{46.97\%}) and F1-score (\textbf{0.4531}). The training history plots (Figure \ref{fig:cifar100_history}) confirm NIRMAL’s stable convergence with minimal oscillations, unlike Adam, which showed more variability.

\section{Conclusion}
This study introduced a novel optimization algorithm, NIRMAL (Novel Integrated Robust Multi-Adaptation Learning), inspired by the movements of chess piece. Our comprehensive comparative analysis against established optimizers, Adam and SGD with Momentum, across four benchmark image classification datasets (MNIST, FashionMNIST, CIFAR-10, and CIFAR-100) highlights NIRMAL's promising capabilities.

NIRMAL consistently demonstrated competitive performance across all evaluated datasets. In particular, on the highly complex CIFAR-100 dataset, NIRMAL outperformed Adam and achieved performance remarkably close to SGD with Momentum, indicating its robustness and effectiveness in challenging deep learning scenarios. The hybrid design of NIRMAL, by combining various optimization strategies, appears to contribute to its improved stability, generalization, and robust convergence properties, especially on more complicated tasks.

Future work will focus on several key areas to further enhance NIRMAL. These include a more extensive hyperparameter optimization study to determine optimal configurations for various network architectures and tasks. Additionally, exploring NIRMAL's applicability to other deep learning tasks beyond image classification, such as natural language processing (NLP), would be valuable. Investigating the theoretical convergence properties and specific contributions of each ``chess piece'' inspired component could also provide deeper insights into the effectiveness of NIRMAL. These avenues of research will further solidify NIRMAL's potential as a versatile and powerful optimizer for the deep learning community.

\bibliographystyle{plainnat}
\bibliography{references}

\begin{thebibliography}{5}
\providecommand{\natexlab}[1]{#1}
\providecommand{\url}[1]{\texttt{#1}}
\expandafter\ifx\csname urlstyle\endcsname\relax
  \providecommand{\doi}[1]{doi: #1}\else
  \providecommand{\doi}{doi: \begingroup \urlstyle{rm}\Url}\fi

\bibitem[Acharya et~al.(2015)Acharya, Paul, and Mitra]{acharya2015sgd}
Apoorv Acharya, Arindam Paul, and Arindam Mitra.
\newblock Comparison of optimization techniques in stochastic gradient descent.
\newblock \emph{International Journal of Advanced Research in Computer and Communication Engineering}, 4\penalty0 (12), 2015.

\bibitem[Kingma and Ba(2015)]{kingma2017adam}
Diederik~P. Kingma and Jimmy Ba.
\newblock Adam: A method for stochastic optimization.
\newblock \emph{International Conference on Learning Representations}, 2015.

\bibitem[Liu et~al.(2020)Liu, Zhou, and Zhou]{liu2020sgd}
Han Liu, Yi~Zhou, and Tianyi Zhou.
\newblock On the convergence of stochastic gradient descent with momentum.
\newblock \emph{International Conference on Machine Learning}, 2020.

\bibitem[Saud and Dhakal(2023)]{saud2023bert}
Arjun~Singh Saud and Ajanta Dhakal.
\newblock Optimizing bert for nepali text classification: The role of stemming and gradient descent optimizers.
\newblock \emph{Asian Journal of Multidisciplinary Research}, 1\penalty0 (1), 2023.
\newblock \doi{10.3126/ajmr.v1i1.82292}.

\bibitem[Soudani and Barhoumi(2019)]{soudani2019image}
Amira Soudani and Walid Barhoumi.
\newblock An image-based segmentation recommender using crowdsourcing and transfer learning for skin lesion extraction.
\newblock \emph{Expert Systems with Applications}, 118:\penalty0 400--410, 2019.
\newblock \doi{10.1016/j.eswa.2018.10.029}.

\end{thebibliography}

\end{document}